\documentclass[aoas,MSNbibl,nameyear,dvips]{arximspdf}

\doi{10.1214/13-AOAS614E} 
\referstodoi{10.1214/12-AOAS614}
\volume{7}
\issue{4}
\pubyear{2013}
\firstpage{1888}
\lastpage{1890}

\begin{document}
\begin{frontmatter}

\title{Discussion of ``Estimating the historical and future
probabilities of large terrorist events'' by~Aaron~Clauset~and~Ryan~Woodard}
\runtitle{Discussion}

\begin{aug}
\author[A]{\fnms{Frederic Paik} \snm{Schoenberg}\corref{}\ead[label=e1]{frederic@stat.ucla.edu}}
\runauthor{F. P. Schoenberg}
\affiliation{University of California, Los Angeles}

\pdftitle{Discussion of ``Estimating the historical and future
probabilities of large terrorist events'' by Aaron Clauset and Ryan Woodard}

\address[A]{Department of Statistics\\
University of California, Los Angeles\\
8125 Math-Science Building\\
Los Angeles, California 90095-1554\\
USA\\
\printead{e1}}
\end{aug}

\received{\smonth{7} \syear{2013}}



\end{frontmatter}

The similarities between terrorist strikes and earthquakes are quite striking.
Clauset and Woodard's Figure~1 shows that the distribution of severity
of terrorist events closely follows the Pareto or power law distribution.
This distribution is also often used to describe the total scalar
seismic moment, or total energy release, of earthquakes [\citet{Uts99}].
It would be interesting to see if the spatial-temporal distribution of
terrorist events appears to be well described by the Epidemic-Type
Aftershock Sequence (ETAS) models of \citet{Oga98}, which are commonly
used to describe earthquakes and which \citet{Mohetal11} recently
used to model the spatial-temporal distribution of violent crimes in
the United States.

In seismology, the {\sl tapered} Pareto distribution has been shown to
offer a somewhat better fit to some earthquake catalogs [\citet{Kag93},
\citet{JacKag99}, Vere-Jones, Robinson and
Yang (\citeyear{VerRobYan01}), \citet{KagSch01}].
Similarly, while several authors have argued that wildfire sizes follow
the pure or truncated Pareto distribution [Strauss, Bednar and Mees (\citeyear{StrBedMee89}),
\citet{Cum01}, Malamud, Millington and
Perry (\citeyear{MalMilPer05})], there is some suggestion that
the tapered Pareto may fit better [Schoenberg, Peng and
Woods (\citeyear{SchPenWoo03}),
\citet{SchPat12}].
This tapered Pareto distribution is somewhat similar to the stretched
exponential considered by Clauset and Woodard, but the tapered Pareto
has a pure exponential tail rather than a stretched exponential tail
and therefore is typically less heavy tailed.
Thus, consideration of the tapered Pareto in the context of terrorism
might possibly lead to the conclusion that very large events, such as
September~11, 2001, are indeed outliers.

Some might object to modeling terrorist strikes, or other events with
human agency, using methods such as those employed by Clauset and
Woodard. Indeed, in addition to the usual statistical problems of
misspecification, overfitting and estimation error, there is the
additional issue that predictions made based on such models can be
self-fulfilling, or self-unfulfilling, or have any of various
intermediate feedback mechanisms based on humans' awareness of these
predictions. However, Clauset and Woodard do not claim to make precise
predictions of future terrorist activity based on their model, and for
the purposes of describing previous terrorist strikes and forecasting
broad trends of future activity, their analysis seems sensible, fair
and honest.

An issue that looms large in seismology is missing data, especially
with regard to historical seismicity or microseismicity [see, e.g., \citet{Kag04}].
In the context of terrorism, it seems unlikely that many large events
are missing from the data set. On the other hand, there must be many
events that could alternately be classified as terrorist events or as
conflicts, battles or even wars, depending on one's political
perspective. Also, both in the case of terrorist events and
earthquakes, it can be extremely difficult and subjective to determine
where one event ends and another begins.
Clauset and Woodard's finding that the simple Pareto distribution
appears to describe the distribution of the severities of terrorist
strikes so well in spite of these subjectivities is remarkable.

While it seems reasonable enough for Clauset and Woodard to treat their
observations as if they were i.i.d., it should be noted that the evidence
in Clauset and Woodard is far from suggesting that these events
actually are i.i.d. Certainly these severities of terrorist attacks could
be nearly Pareto distributed whether the events are i.i.d. or not. Indeed,
the notion that these events are i.i.d. seems to contrast with the obvious
nonstationarity in Clauset and Woodard's Figure~3. In seismology,
earthquakes appear very obviously to arrive in clusters, both spatially
and temporally. A large earthquake typically has many moderate to large
aftershocks in its spatial-temporal vicinity, for example. One would
anticipate terrorist events to also exhibit clustering, though perhaps
not as intense as in the seismological context. It would be interesting
to consider and attempt formally to test whether the sizes of terrorist
events may be i.i.d. and separable, in the sense of Cressie (\citeyear{Cre91}), from
the locations and times of the events, as in the ETAS model; tests for
this purpose were developed, for example, in \citet{Sch04},
Assuncao and Maia (\citeyear{AssMai07}) and \citet{ChaSch11}.

%



\printaddresses

\end{document}